\documentclass[showpacs,prl,onecolumn,aps,superscriptaddress,preprintnumbers,nofootinbib]{revtex4}
\usepackage{amsmath,amssymb}
\usepackage{epsfig}
\usepackage{graphicx}
\usepackage{mathrsfs}
\usepackage{amsmath}
\usepackage{amsfonts}
\usepackage{epstopdf}
\usepackage{color}
\def\slashchar#1{\setbox0=\hbox{$#1$}     		
   \dimen0=\wd0                                 	
   \setbox1=\hbox{/} \dimen1=\wd1               	
   \ifdim\dimen0>\dimen1                        	
      \rlap{\hbox to \dimen0{\hfil/\hfil}}      	
      #1                                        	
   \else                                        	
      \rlap{\hbox to \dimen1{\hfil$#1$\hfil}}   	
      /                                         	
   \fi}

\newcommand{\be}{\begin{equation}}
\newcommand{\ee}{\end{equation}}
\newcommand{\bea}{\begin{eqnarray}}
\newcommand{\eea}{\end{eqnarray}}
\newcommand{\ba}{\begin{array}}
\newcommand{\ea}{\end{array}}

\def\eq#1{{Eq.~(\ref{#1})}}
\def\fig#1{{Fig.~\ref{#1}}}

\newcommand{\Lb}{\left(}
\newcommand{\Rb}{\right)}
\newcommand{\h}{\frac{1}{2}}

\begin{document}

\title{Deep inelastic scattering as a probe of entanglement:\\ confronting experimental data}

\author{Dmitri E. Kharzeev}
\email{Dmitri.Kharzeev@stonybrook.edu}
\affiliation{Center for Nuclear Theory, Department of Physics and Astronomy, Stony Brook University, New York 11794-3800, USA}
\affiliation{Department of Physics and RIKEN-BNL Research Center, \\
Brookhaven National Laboratory, Upton, New York 11973-5000, USA}
\author{Eugene Levin}
\email{leving@tauex.tau.ac.il, eugeny.levin@usm.cl}
\affiliation{Department of Particle Physics, School of Physics and Astronomy,\\
Tel Aviv University, Tel Aviv, 69978, Israel}
\affiliation{Departamento de F\'\i sica,
Universidad T$\acute{e}$cnica Federico Santa Mar\'\i a   and
Centro Cient\'\i fico-Tecnol$\acute{o}$gico de Valpara\'\i so,
Casilla 110-V,  Valparaiso, Chile}

\date{\today}


\begin{abstract}
Parton distributions can be defined in terms of the entropy of entanglement between the spatial region probed by deep inelastic scattering (DIS) and the rest of the proton. For very small $x$,  the proton becomes a maximally entangled state. This approach leads to a simple relation $S = \ln N $  between the average number $N$ of color-singlet dipoles in the proton wave function and the entropy of the produced hadronic state $S$. At small $x$, the multiplicity of dipoles is given by the gluon structure function, $N = x G(x,Q^2)$. 
Recently, the H1 Collaboration analyzed the entropy of the produced hadronic state in DIS, and studied its relation to the gluon structure function; poor agreement with the predicted relation was found.   In this letter we argue that a more accurate account of 
the number of color-singlet dipoles in the kinematics of H1 experiment (where hadrons are detected in the current fragmentation region) is given not by $xG(x,Q^2)$ but by the sea quark structure function $x\Sigma(x,Q^2)$. 
Sea quarks originate from the splitting of gluons, so at small $x$  $x\Sigma(x,Q^2)\,\sim\, xG(x,Q^2)$, but in the current fragmentation region this proportionality is distorted by the contribution of the quark-antiquark pair produced by the virtual photon splitting. In addition, the multiplicity of color-singlet dipoles in the current fragmentation region is quite small, and one needs to include $\sim 1/N$ corrections to $S= \ln N$ asymptotic formula. Taking both of these modifications into account, we find that the data from the H1 Collaboration in fact agree well with the prediction based on entanglement.

\end{abstract}
\maketitle

In our paper \cite{KHLE} (see also \cite{Tu:2019ouv,GOLE}) we computed 
 the von Neumann entropy of the system of partons resolved by deep inelastic scattering (DIS) at a given Bjorken $x$ and momentum transfer $q^2 = - Q^2$.  We then proposed to interpret it as the entropy of entanglement between the spatial region probed by deep inelastic scattering and the rest of the proton. We found that 
in the small $x$, large rapidity $Y$ regime, all partonic micro-states have equal probabilities -- the proton is composed by an exponentially large number $N$ of micro-states that occur with equal and  small probabilities $1/N$. 
This yields a simple relation between the entanglement entropy and the multiplicity of partons (dominated by gluons at small $x$):
\be \label{S}
S = \ln[N]\,\
 \ee
 where $N( x, Q^2)$ is an average  number of color-singlet dipoles. In the region of small $x$  the gluons dominate and $N \,\simeq \,x G(x, Q^2)$ where $xG(x, Q^2)$ is the gluon structure function\footnote{Note that this relation is a quantum analog of the Boltzmann formula underlying statistical physics.}. 
 Assuming that the multiplicity of produced hadrons is proportional to the multiplicity of color-singlet dipoles  (``local parton-hadron duality" \cite{LHPD,MULIB}), eq. (\ref{S}) imposes a relation between the parton structure function (extracted from the inclusive cross section of DIS) and the entropy of produced hadrons; this relation can be directly tested in experiment. The comparison to the experimental data on hadron multiplicity distributions from CMS Collaboration at the LHC provided encouraging results  \cite{KHLE,Tu:2019ouv}.
 \vskip0.3cm
 
\begin{figure}[t]
\begin{center}
\begin{tabular}{cc}
\includegraphics[width=0.5\textwidth]{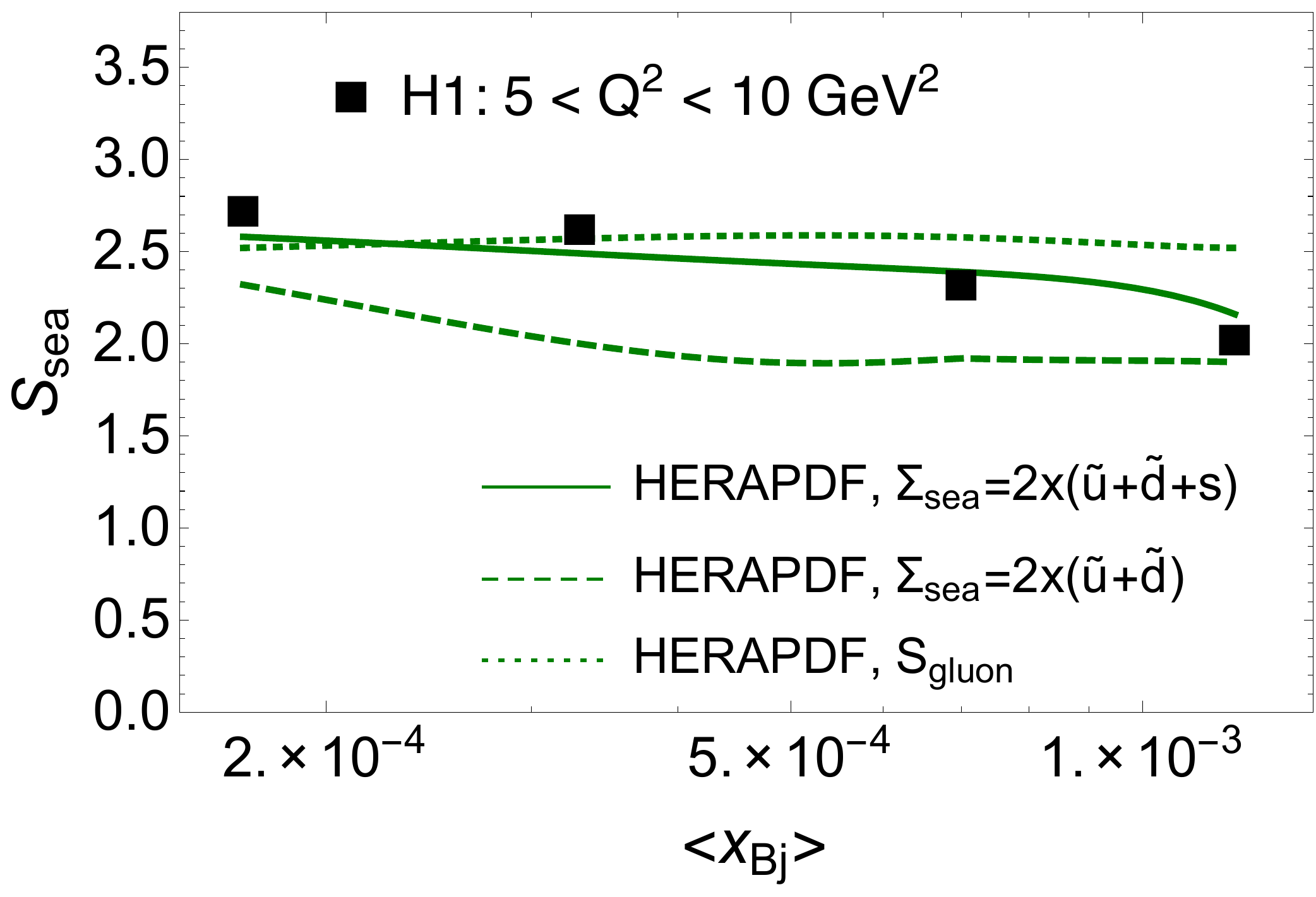}&\includegraphics[width=0.5\textwidth]{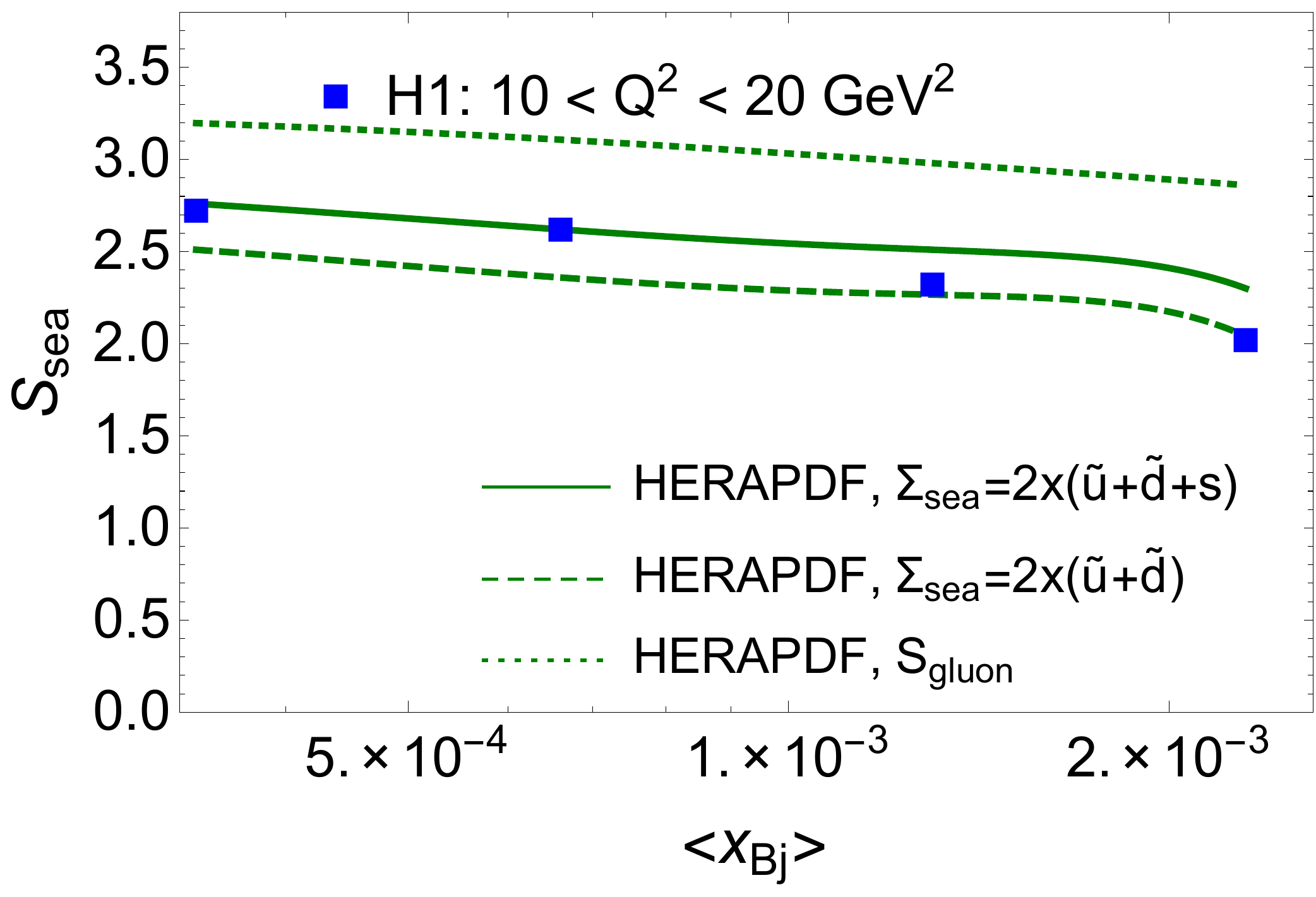}\\
\includegraphics[width=0.5\textwidth]{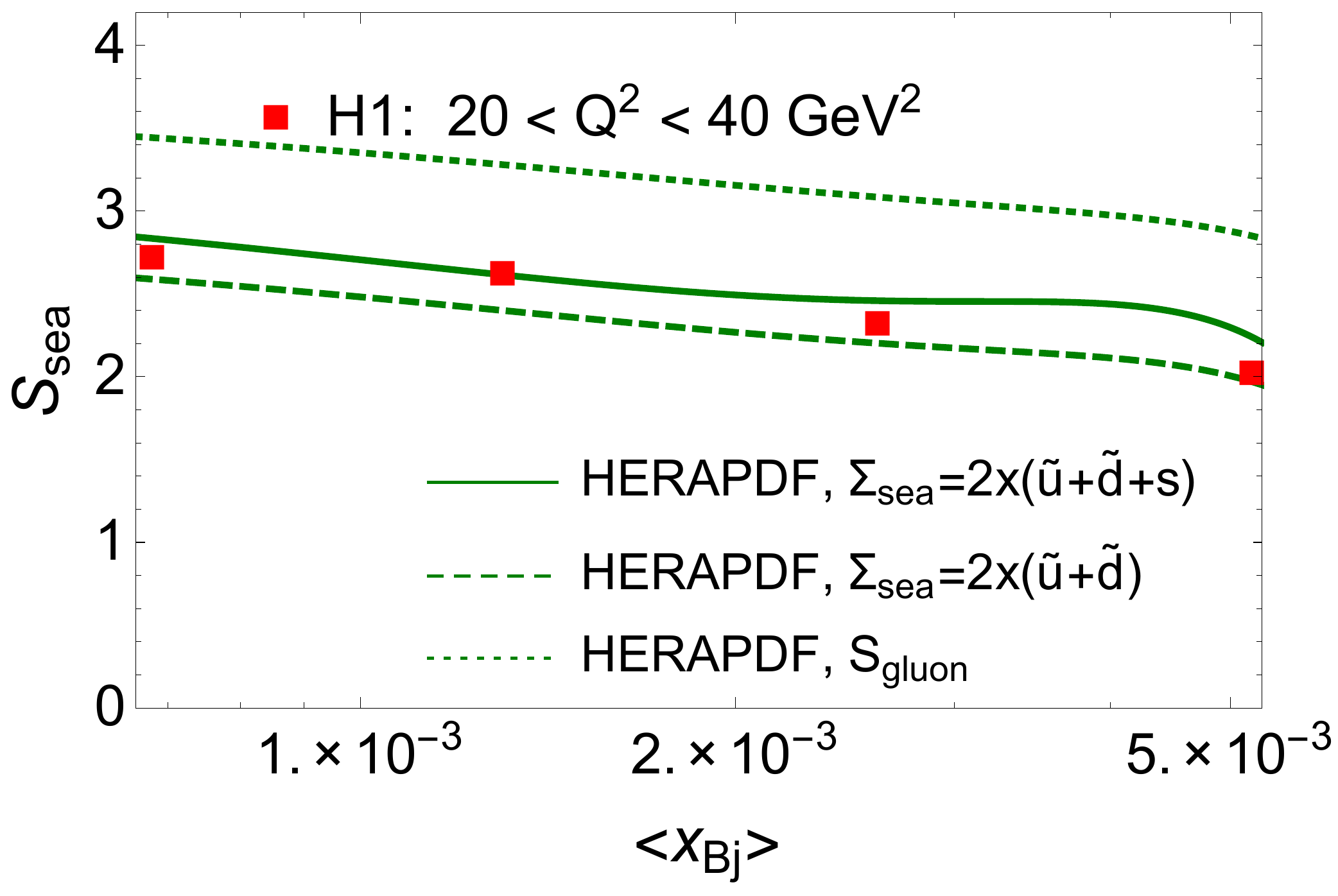}&\includegraphics[width=0.5\textwidth]{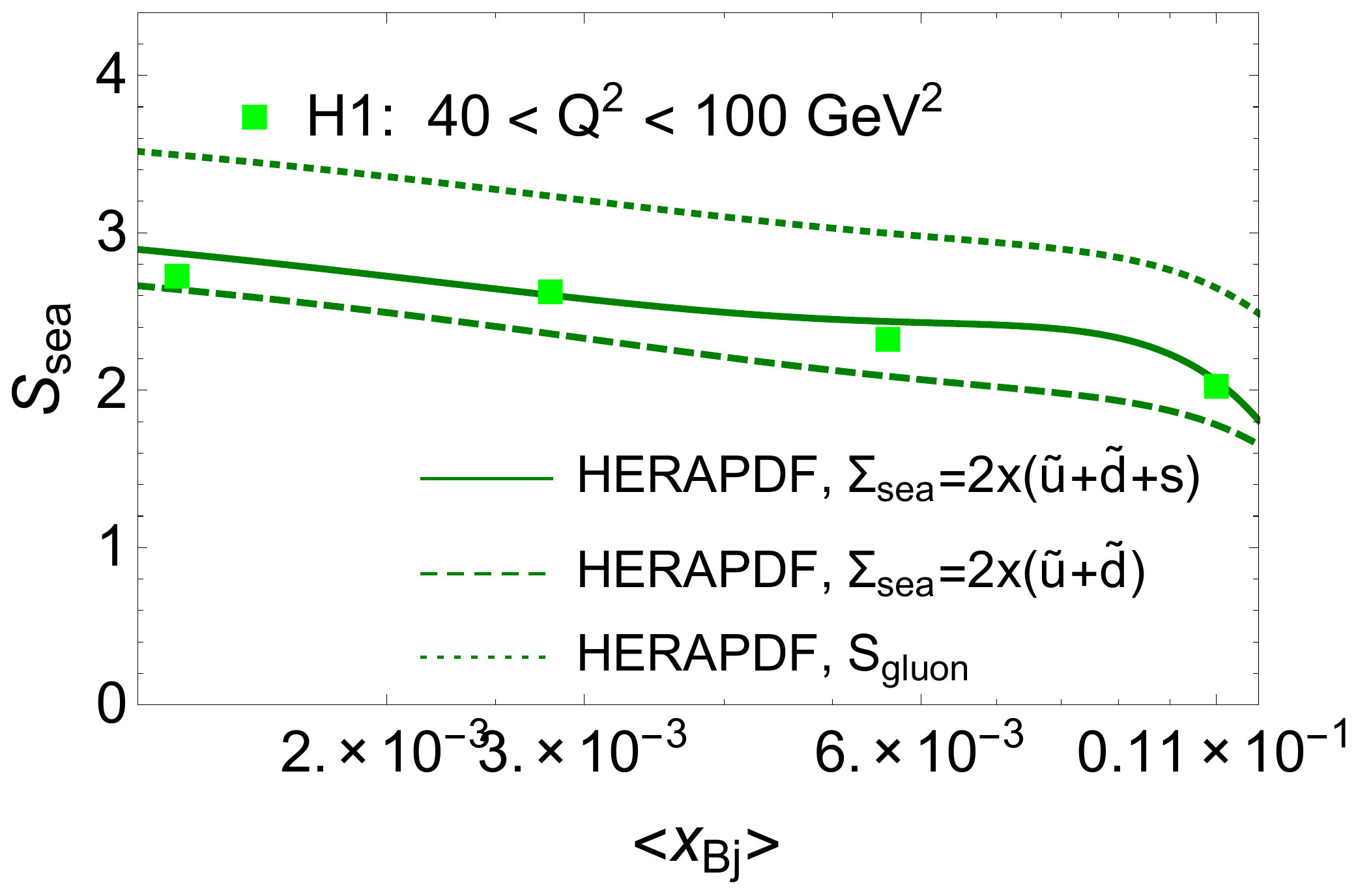}\\
\end{tabular}\end{center}
\caption{Comparison of 
the experimental data of the H1 collaboration \cite{H1EE} on the entropy of produced hadrons in DIS \cite{H1EE} with our theoretical predictions, for which we use the sea quark distributions from the NNLO fit\cite{H1ZEUS,KATARZYNA} to the combined H1-ZEUS data. }
\label{ex}
\end{figure}

\begin{figure}[h]
\begin{center}
\includegraphics[width=0.5\textwidth]{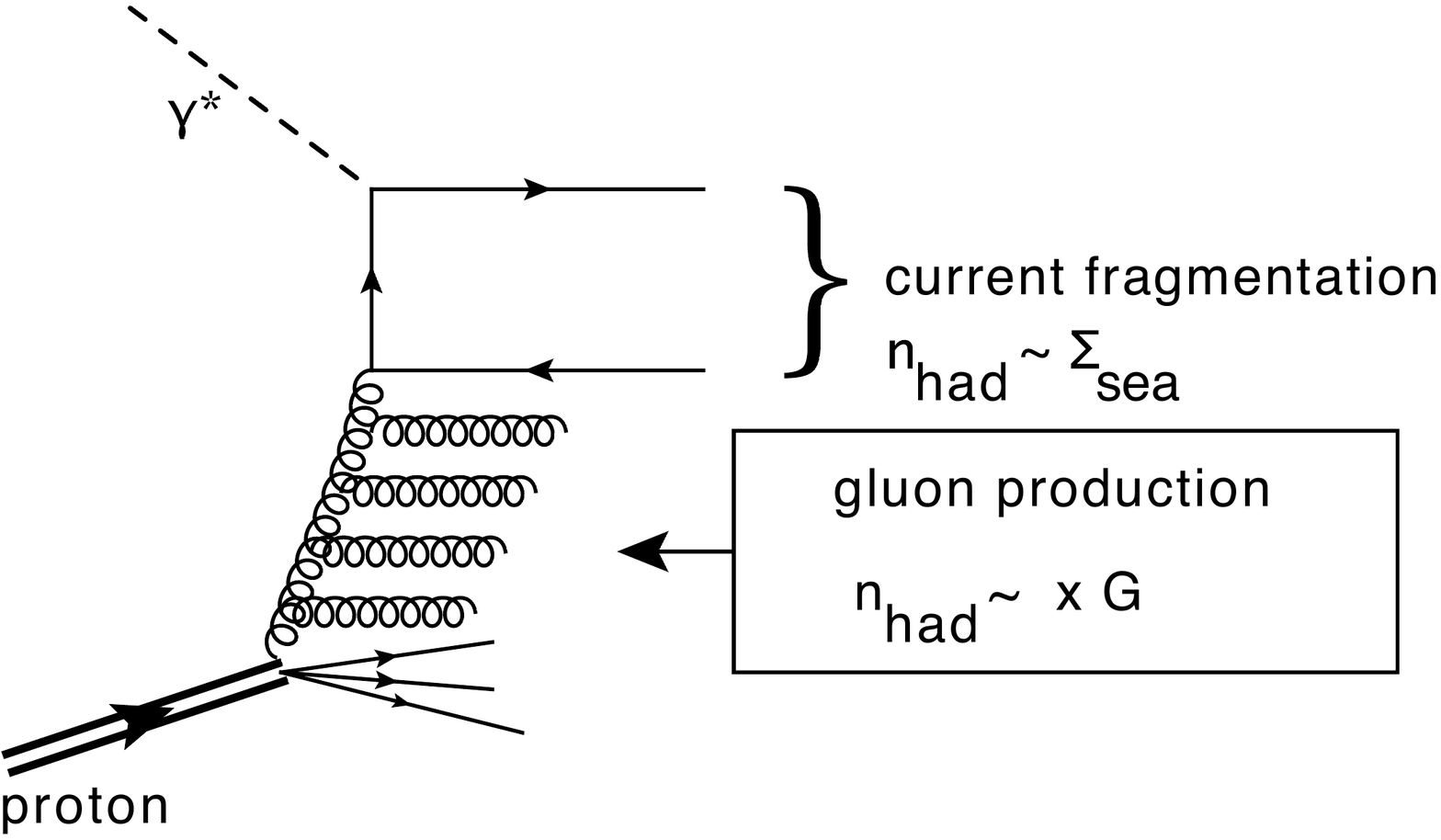}
\end{center}
\vspace{-3.5cm}
\caption{ DIS at small $x$.  }
\label{pc}
\end{figure}

 However, recent dedicated experimental analysis performed by the H1 collaboration \cite{H1EE} shows a disagreement with \eq{S} if one assumes $N \,= \,x G(x, Q^2)$ (see Fig. 12 in  Ref. \cite{H1EE} and the dotted curves in our \fig{ex}). In this letter we demonstrate that the H1 data in fact are in a good agreement with our approach, once two important effects are taken into account. Both of them are implied by the kinematics of the H1 measurements that are performed in the current fragmentation region (see also Refs.\cite{H1EE,H1MULT1,H1MULT2,ZEUSMULT}), and thus at moderate values of Bjorken $x$.
 First, since the experimental hadron multiplicities are not large, we need to take into account corrections of the order of $1/N$ to \eq{S}. Second, because the H1 data are concentrated in the current fragmentation region  and not at very small $x$,   we need to reconsider our claim in Ref.\cite{KHLE} that the multiplicity of the color-singlet dipoles is equal to $xG(x, Q^2)$.  Indeed, the hadrons produced in the current fragmentation  region of DIS originate from the hadronization of the struck quark  (a constituent of the color-singlet dipole) and the multiplicity of color-singlet dipoles is thus determined by the sea quark structure function, see Fig. \ref{pc}. Therefore the correct relation between the number of dipoles and  the experimentally measured entropy of hadrons (valid for large $N$) is
 \be \label{SSI}
 S_{\rm dipoles}\,=\,\ln (x \Sigma(x,Q^2))\,=\,S_{\rm hadron}.
 \ee
  The sea quark and gluon distributions are related to each other at small $x$ by  (see Fig. \ref{pc})
 \be
 x \Sigma(x,Q^2 )\,\,=\,\,C(\alpha_s \ln Q^2,\alpha_s,\alpha_s\ln(1/x))\,\,x G( x, Q^2) ,
 \ee
 where the function $C(\alpha_s \ln Q^2,\alpha_s,\alpha_s\ln(1/x))$ describes the splitting of the virtual photon into the quark-antiquark pair.   In the region of very small $x$, as we will now show, $C(\ln Q^2,\alpha_s,\alpha_s\ln(1/x)) \,\to {\rm Const}$, and so the sea quark distribution is proportional to the gluon one. Therefore, with a logarithmic accuracy at very small $x$ we recover our original relation
 \be\label{orig}
 S_{\rm hadron} \,=\, \ln (x G(x,Q^2)) .
 \ee
 To show that $C(\ln Q^2,\alpha_s,\alpha_s\ln(1/x)) \,\to {\rm Const}$ at very small $x$, let us 
 use the leading order  DGLAP evolution equation \cite{DGLAP} that gives for the sea quark structure function  (see Fig. \ref{pc})
  \be \label{DGLAP}
   x \Sigma(x,Q^2 )  \,\,=\,\,\frac{ C_F \,\alpha_s}{2\,\pi} \int^\xi_0 d \xi' \int^1_x d z\,P_{qG}\Lb z\Rb \Lb \frac{x}{z} G\Lb \frac{x}{z}, \xi'\Rb\Rb~~~~\mbox{with}~~P_{qG}\Lb z\Rb = \frac{ 1 + (1-z)^2}{z} , \ee 
 where  $\xi = \ln Q^2$.
 At small values of $x$, the gluon structure function takes the form (see Ref.\cite{KOLEB} for a review) 
  \be \label{GMLN}
  x G( x, \xi)\,\,=\,\,\int^{\epsilon + i \infty}_{\epsilon - i \infty} \frac{d \gamma}{ 2\,\pi\,i} \Lb \frac{1}{x}\Rb^{ \alpha_s \frac{N_c}{\pi} \chi\Lb \gamma\Rb } e^{\gamma\,\xi} g\Lb \gamma\Rb_{in} 
  \ee
  where $N_c $ is the number of colors, $C_F = \Lb N^2_c -1\Rb/2 N_c$, and the BFKL kernel $\chi\Lb \gamma\Rb $ has the form:
  \be \label{CHI}
  \chi\Lb \gamma \Rb  =  2 \psi\Lb 1\Rb \,-\,\psi\Lb \gamma\Rb\,-\,\psi\Lb 1 - \gamma\Rb  \xrightarrow{\gamma \to \frac{1}{2}}\,\underbrace{ 4 \ln 2}_{\omega_0} \,\,+\,\, \underbrace{14 \zeta\Lb 3\Rb}_{D} \Lb \gamma - \h\Rb^2 \,\,\mbox{in}\,\,\,\,\mbox{diffusion approximation};  
  \ee
  $g\Lb \gamma\Rb_{in} $ should be calculated from the initial conditions.
  Plugging \eq{GMLN} in \eq{DGLAP} we can take the integrals over $z$ and $\xi'$.  Taking the integral over 
  $\gamma$ in the diffusion approximation, and using the method of steepest descent , we obtain the following result:
 \be  \label {XSI}
   x \Sigma(x,Q^2 ) \,\,=\,\, \frac{2\,C_F}{\omega_0\,N_c}  x G( x, Q^2) \,\,=\,\,{\rm Const} \, x G( x, Q^2) 
      \ee
 We evaluated \eq{XSI}  for small $\alpha_s$ neglecting terms in $C(\alpha_s \ln Q^2,\alpha_s,\alpha_s\ln(1/x))$  proportional to $x$.  Note that ${\rm Const}$ does not depend on $\alpha_s$ and numerically is about 0.3. Actually, the next -to-leading order correction increases the value of this ${\rm Const} $ and, as we can see  from Fig. \ref{ex} (panel with $5\, \leq Q^2 \,\leq \,10\,{\rm GeV}^2$), ${\rm Const} \sim 1 $  in the NNLO fit for  the region of smallest $x$. However, away from the region of very small $x$ and in the kinematics of H1 measurement, the proportionality (\ref{XSI}) is violated, and one should use the relation (\ref{SSI}) instead of (\ref{orig}).




\begin{figure}[h]
\begin{center}
\includegraphics[width=0.4\textwidth]{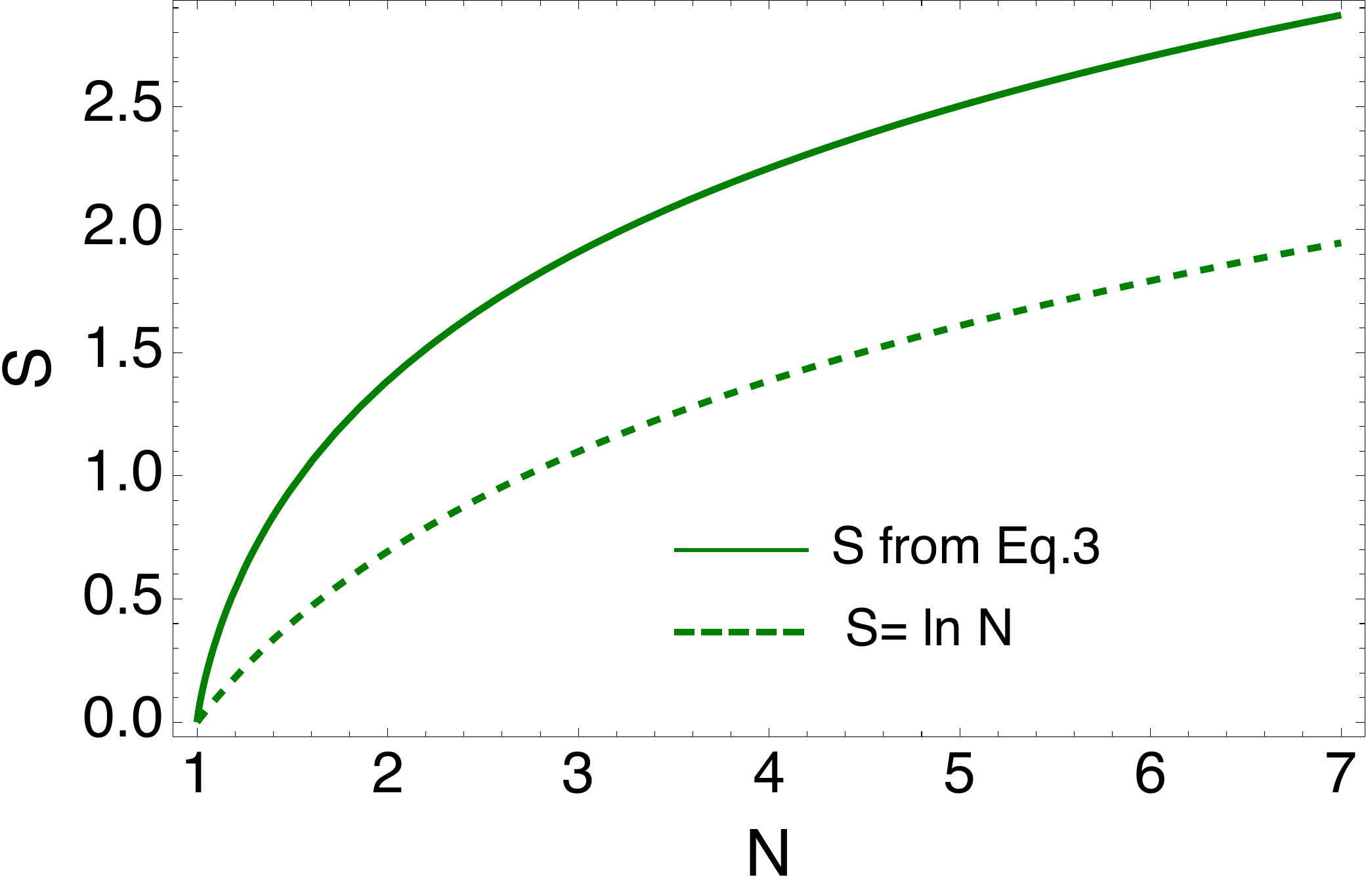}
\end{center}
\vspace{-0.7cm}
\caption{ Entropy versus multiplicity N from \eq{S} and \eq{SS}.  }
\label{ss}
\end{figure}

 \vskip0.3cm
In addition, away from small $x$, where the multiplicity of color-singlet dipoles $N$ is not large, one should take into account $1/N$ corrections to (\ref{orig}). 
In Refs.\cite{KHLE,GOLE} it is shown that in QCD cascade the multiplicity distribution has the following form:

\be \label{PN}
p_n\Lb N \Rb\,\,=\,\,\frac{1}{N}\Lb 1\,-\, \frac{1}{N}\Rb^{n - 1} 
\ee
where $N$ is the average multiplicity of color-singlet dipoles. The distribution (\ref{PN}) leads to the following von Neumann entropy:
\be \label{SS}
S\,=\,-\sum p_n \ln p_n \,\,=\,\,\ln(N - 1)\,\,+\,\,N \ln\Lb 1 + \frac{1}{N - 1}\Rb
\ee
One can see that at large $N$ we obtain $S \simeq \ln N$, but corrections are sizable when $N \leq 10$ (see \fig{ss}). It should be noted that the distribution of \eq{PN} describes quite well the experimental hadron multiplicity distributions in proton-proton collisions (see Refs. \cite{KHLE,Tu:2019ouv,GOLE}).

For comparison with the H1 experimental data \cite{H1EE}(see \fig{ex}), we first assume, following \cite{KHLE}, that the hadron multiplicity is equal to the number of color-singlet dipoles. This assumption is 
  based on ``parton liberation" picture \cite{MULIB} and on the "local
 parton-hadron duality" \cite{LHPD}.    For  sea quark and  gluon structure functions   in \fig{ex} we use  NNLO fit \cite{H1ZEUS,KATARZYNA} to the combined H1 and ZEUS data.


\vskip0.3cm

One can see that our approach in fact describes the H1 data quite well --  this is the first test of the 
relation between entanglement and the parton model in DIS enabled by the H1 analysis. We stress that once the data in the target fragmentation region  at smaller value of $x$ 
becomes available at the Electron-Ion Collider, one should be able to use $xG(x,Q^2)$ in the relation (\ref{S}), as  it has been done in Refs. \cite{KHLE,Tu:2019ouv,Baker:2017wtt,GOLE,GOLE1}. However, the general formula  is given by \eq{SSI} which at small $x$ reduces to  $S \,=\,\ln( xG(x,Q^2))$ since $x\Sigma(x,Q^2) \,\to\,xG(x,Q^2)$.

\vskip0.5cm
   
 {\bf Acknowledgements:}   We thank Kong Tu, Thomas Ullrich and our colleagues at BNL, Stony Brook University, Tel Aviv University and UTFSM for
 stimulating discussions. We are very grateful to Aharon Levy and Katarzyna  Wichmann for help in finding and extracting the parton distributions in the NNLO fit to HERA data. This work was supported in part by the U.S. Department of Energy under Contracts No. DE-FG88ER40388 and DE-SC0012704, BSF grant   2012124, ANID PIA/APOYO AFB180002 (Chile) and  Fondecyt (Chile) grants  
 1180118.

    \end{document}